\documentclass[10pt,journal,compsoc]{IEEEtran}

\ifCLASSOPTIONcompsoc
  \usepackage[nocompress]{cite}
\else
  \usepackage{cite}
\fi
\ifCLASSINFOpdf
  \usepackage[pdftex]{graphicx}
\else
\fi
\usepackage{amsmath}
\usepackage{algorithmic}
\usepackage[ruled,linesnumbered]{algorithm2e}

\usepackage{array}

\ifCLASSOPTIONcompsoc
  \usepackage[caption=false,font=footnotesize,labelfont=sf,textfont=sf]{subfig}
\else
  \usepackage[caption=false,font=footnotesize]{subfig}
\fi
\usepackage{url}
\begin{document}
%
\title{Acceleration for Timing-Aware Gate-Level Logic \\ Simulation with One-Pass GPU Parallelism}
%
%
%
%

\author{
    Weijie~Fang,
    Yanggeng~Fu,
    Jiaquan~Gao,
    Longkun~Guo,
    Gregory~Gutin,
    and~Xiaoyan~Zhang
\IEEEcompsocitemizethanks{
\IEEEcompsocthanksitem W. Fang and L. Guo are with the School of Mathematics and Statistics, Fuzhou University, Fuzhou, Fujian, China, 350108.\protect\\
E-mail: longkun.guo@gmail.com
\IEEEcompsocthanksitem Y. Fu is with the College of Computer and Data Science/College of Software, Fuzhou University, Fuzhou, Fujian, China, 350108.
\IEEEcompsocthanksitem J. Gao is with the School of Computer and Electronic Information, Nanjing Normal University, Nanjing, Jiangsu, China, 210023.
\IEEEcompsocthanksitem G. Gutin is with the Department of Computer Science, Royal Holloway, University of London, Egham, Surrey, UK, TW20 0EX.
\IEEEcompsocthanksitem X. Zhang is with the School of Mathematical Science and the Institute of Mathematics, Nanjing Normal University, Nanjing, Jiangsu, China, 210023.
}
\thanks{Manuscript received April 19, 2005; revised August 26, 2015.}
}

\IEEEtitleabstractindextext{%
\begin{abstract}
Witnessing the advancing scale and complexity of chip design and benefiting from high-performance computation technologies, the simulation of Very Large Scale Integration (VLSI) Circuits imposes an increasing requirement for acceleration through parallel computing with GPU devices. However, the conventional parallel strategies do not fully align with modern GPU abilities, leading to new challenges in the parallelism of VLSI simulation when using GPU, despite some previous successful  demonstrations of significant acceleration. In this paper,  we propose a novel approach to accelerate 4-value logic timing-aware gate-level logic simulation using waveform-based GPU parallelism. Our approach utilizes a new strategy that can effectively handle the dependency between tasks during the parallelism, reducing the synchronization requirement between CPU and GPU when parallelizing the simulation on combinational circuits. This approach requires only one round of data transfer and hence achieves one-pass parallelism. Moreover, to overcome the  difficulty within the adoption of our strategy in GPU devices, we design a series of data structures and tune them to dynamically allocate and store new-generated output with uncertain scale. Finally,  experiments are carried out on industrial-scale open-source benchmarks to demonstrate the performance gain of our approach compared to several state-of-the-art  baselines. 
\end{abstract}

}

\maketitle

\IEEEdisplaynontitleabstractindextext

%
\IEEEpeerreviewmaketitle

\ifCLASSOPTIONcompsoc
\IEEEraisesectionheading{\section{Introduction}\label{sec:introduction}}
\else
\section{Introduction}
\label{sec:introduction}
\fi

%
%
%
%
\IEEEPARstart{T}{iming-aware} gate-level logic simulation has become a vital component in many existing tasks of VLSI verification. If the correct behavior of a circuit is provided by an Automatic Test Pattern Generation (ATPG) or some other previous tasks in the problem of timing-aware gate-level logic simulation, the problem is called re-simulation since the provided waveforms can be obtained by a previous Register Transfer Level (RTL) simulation. In this case, the complete waveforms of several pins that consist of consecutive transitions in a certain duration will be given. These waveforms usually belong to special cell pins in the netlist such as primary inputs, register output, or sequential signals. The objective of timing-aware gate-level logic simulation is to obtain all waveforms of output pins in the netlist based on the provided waveforms of the input pins. Logic simulation usually appears in applications that desire massive and complex verification such as the detection of hardware trojans \cite{xiao2016hardware}, the generation of Design-For-Test (DFT) patterns \cite{press2008ic}, and fault simulation or power analysis for netlist designs. Compared with RTL simulation, timing-aware gate-level logic simulation is usually more time-consuming and computation-intensive, and may take hours or days to accurately simulate the result in a large-scale System on Chip (SoC). Therefore, gate-level logic simulation raises a more demanding challenge for being accelerated when using highly-parallel computing systems such as Graphics Processing Units (GPU). GPU accelerated techniques have been significantly prompted over the past years. Meanwhile, many existing works on accelerated approaches have proved their success in the combination of GPU parallelism and VLSI verification, despite most of them being concerned with 2-value (i.e., ``0'' and ``1'') logic simulation or analog circuit simulation.

Verilog Hardware Description Language (VHDL) \cite{thomas2002verilog}, which uses 4-value logic in simulation, has become the current universal hardware description language in the fields of Electronic Engineering. Different from the well-known 2-value logic, 4-value logic uses values ``X'' and ``Z'' besides ``0'' and ``1'', by which the simulation can be more accurate for industrial applications while being more difficult to be processed. In this language, ``X'' is an abstract symbol that describes an unassigned signal, or more typically, an unknown signal whose electrical level is higher than ``0'' and lower than ``1'', which may go into effect as either ``0'' or ``1'' in practice. And ``Z'' denotes the high impedance state, an existent output signal in the real world that does not influence its successor circuits.

\subsection{Relate Works}

Perinkulam \cite{perinkulam2007logic} investigated the implementation and bottlenecks of GPU-based approaches compared with the CPU-based ones and demonstrated that GPU could be helpful for the tasks of logic simulation.
Gulati et al. \cite{gulati2008towards} implemented a thread-level parallel approach for fault simulation based on the large memory bandwidth of GPUs and Kochte et al. \cite{kochte2010efficient} introduced the parallel-pattern single-fault propagation paradigm into the fault simulation algorithm.
Chatterjee et al. \cite{chatterjee2011high} achieved up to 62x acceleration of RTL simulation on an NVIDIA 8000GT GPU compared to the baseline obtained by a 3.4 GHz Pentium 4 processor, which benefits from but is also restricted to the utilization of GPU shared memory.
Qian et al. \cite{qian2011accelerating} translated the VHDL RTL description into GPU source code and adopted a CMB-based parallel simulation protocol aiming at more effective parallelism.
Zhang et al. \cite{zhang2011logic} addressed an accelerated VLSI logic simulation approach on GPU with an adaptable partition strategy and reached up to 21x speedup compared with the single-core computation.
Meraji et al. \cite{meraji2012optimizing} employed a dynamic load balancing strategy incorporated with a bounded window restraining the optimism of time warp, which resulted in up to 60\% time cost improvement on open-source Sparc, Leon, and two Viterbi decoders designs.
Chen et al. \cite{chen2015sparse} developed a hybrid parallel approach to address the bottleneck at sparse matrix LU factorization in circuit simulators based on simulation programs with integrated circuit emphasis, which combines task-level and data-level parallelism and optimizes for work partitioning, number of active thread groups and memory access patterns based on the GPU architecture.
Chhabria et al. \cite{chhabria2022crosstalk} incorporated GPU-accelerated dynamic gate-level simulation and machine learning, which can tackle the disturbance from false aggressors and predict the delta delays of identified crosstalk-critical nets.
Zhang et al. \cite{zhang2022GATSPI} presented a GPU-accelerated gate-level simulation approach for power improvement, which achieved ultra-fast power estimation on industry-sized ASIC designs with millions of logic gates and results in up to 7412x speedup on 2-value logic re-simulation compared with a single-core commercial simulator if applying the approach on a multi-GPU parallel system.

Xu et al. \cite{xu2004predicting} established a model to predict the performance of synchronization in parallel discrete event simulation and proved its accuracy on several large-scale ISCAS logic circuit simulations. In the work, the authors also discussed the factors that help to improve the performance of synchronous parallelism, indicating that communication implementation and partitioning strategy are vital to parallel efficiency.
Sen et al. \cite{sen2010parallel} leveraged the property of a directed acyclic graph (DAG)\footnote{Many properties of DAGs are shown in the survey chapter \cite{gutin2018classes} on DAGs.} to partition the netlist and then proposed a cycle-based simulation approach.
Zhu et al. \cite{zhu2011massively} attempted to reduce the communication between CPU and GPU using a memory paging structure with a data-mapping strategy and received a 270x acceleration compared with the CPU simulator, but it is different in essence from the method in this paper.
Chatterjee et al. \cite{chatterjee2009event} and Holst et al. \cite{holst2015high} contributed to the event-based gate-level simulation approaches, and the former founded this methodology while the latter achieved up to 1000x speedup on the re-simulation of a 2-value logic circuit that supports finite kinds of logic gates.
Zeng et al. \cite{zeng2021accelerate} proposed a 2-dimensional event-based parallelism re-simulation approach based on grouping gates and splitting events and further compressing the states to reduce the communication overhead, which is the only recent work in 4-value logic simulation on DAG netlists\footnote{The abstract graph of the netlist in re-simulation can also be regarded as a DAG since the incoming edges of given pseudo-primary inputs are useless and can be removed, thus making the graph acyclic.} to the best of authors’ knowledge.

\begin{figure}[t]
    \centering
    \includegraphics[width=3.2in]{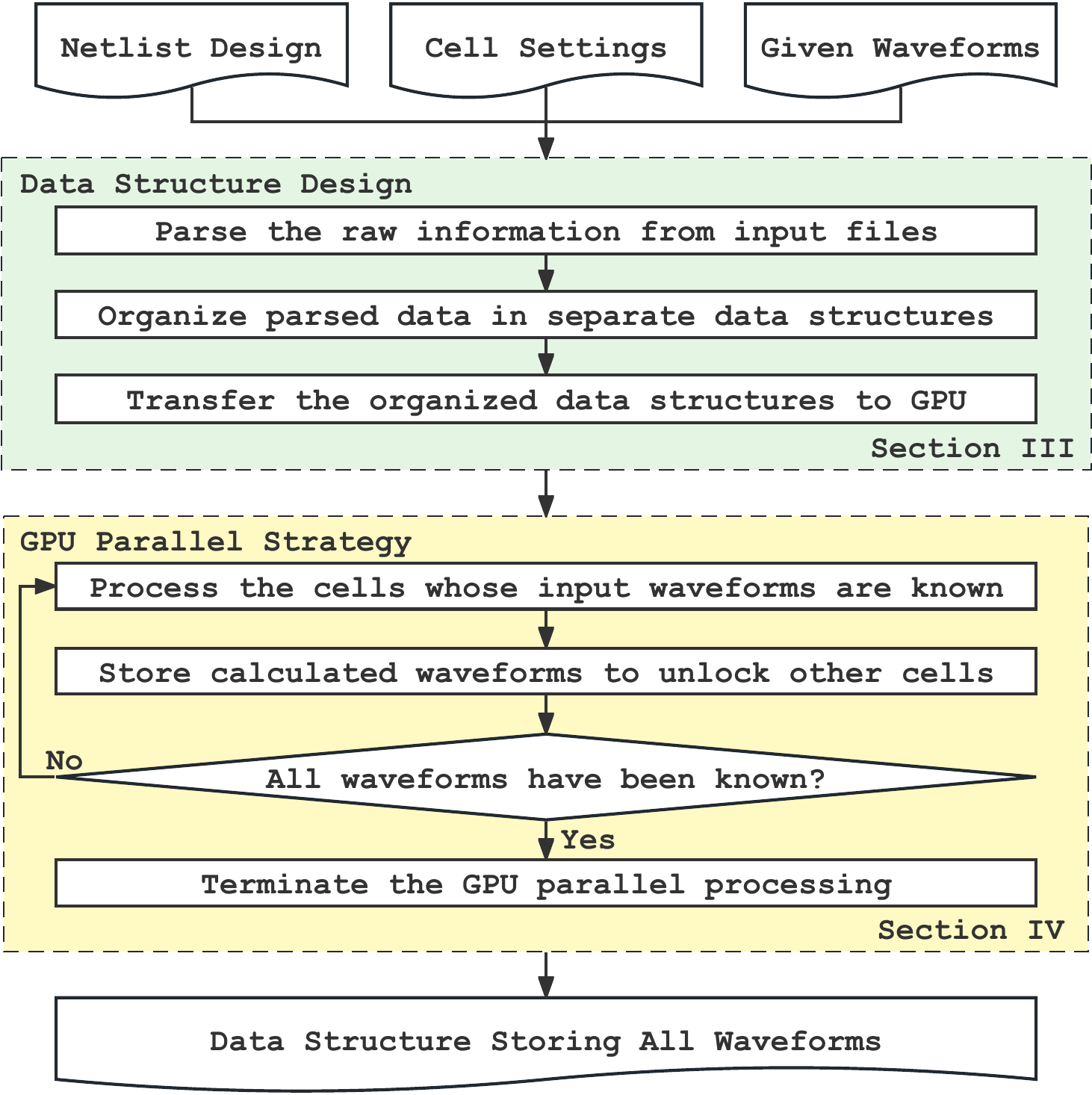}
    \caption{Diagram of the proposed approach.}
    \label{diagram}
\end{figure}

\subsection{Motivation}

As demonstrated by Xu et al. \cite{xu2004predicting}, synchronization consumes a considerable portion of runtime in synchronous parallelism. Existing simulation approaches with GPU parallelism present an open problem concerning the reduction of frequent synchronization between CPU and GPU, which results from the inefficient utilization of advanced GPU memory. In contrast, SOTA GPUs have already acquired up to 80GB memory \cite{A100}, which can be further expanded through tools such as NVLink and NVSwitch \cite{hopper}.
Thus, it naturally arises a challenge to reduce synchronization frequency via better utilizing the large GPU memory. On the other hand, the parallelism of 4-value logic simulation has broad applications in the industry but is only addressed by a few existing works in literature.
Combining the above-mentioned challenges and tasks, we are particularly interested in one-pass GPU parallelism for 4-value logic simulation, where one-pass GPU parallelism means only one round of data transfer is needed for the entire computing task and optimally minimizes the cost of synchronization.


\subsection{Our Contributions}

In this paper, we propose an approach for accelerating gate-level logic simulation tasks on DAG netlists with one-pass GPU parallelism,   including acceleration of  logic simulation on combinational circuits and re-simulation. The contributions can be summarized as follows:
\begin{itemize}
    \item We design novel data structures to support our one-pass parallelism. The devised data structures significantly reduce the frequency of memory reallocation during the parallelism, resulting in no frequent synchronization between GPU and CPU and comparing favorably to the existing event-based method.
    \item We convert the problem into a model different from the previous ones on DAG and propose a waveform-based GPU parallel strategy supporting 4-value logic. GPU code is employed to control the processing of dependent tasks in parallelism and produce output that contains no invalid states as the input for the successor netlist.
    \item Extensive experiments were carried out on open-source benchmarks, demonstrating that our approach outperforms the previous SOTA on average. In particular, it is shown that our approach achieves a remarkably higher utilization of GPU computing resources when the time costs of the parallel tasks significantly differ.
\end{itemize}

\subsection{Algorithmic Flow}

The key idea of our approach is first to represent the given netlist using specifically designed data structures and then transfer them concurrently into the memory of GPU devices for computing in parallel. Within the duration of parallel computing, we use GPU threads to calculate waveforms, unlock new cells, and maintain outputs, without any communication with the CPU. After completing the work of all GPU threads, the result will be transferred back to CPU, resulting in one round of data transfer in total.
Fig. \ref{diagram} shows the diagram of our algorithmic flow.

\subsection{Organization}

The remainder of this paper is organized as follows:
Section 2 states the preliminary and hardcore mission of 4-value logic timing-aware gate-level logic simulation;
Section 3 illustrates the design of vital data structure serving for the one-pass parallelism;
Section 4 explains how the proposed waveform-based GPU accelerated method operates;
Section 5 shows the result of experiments on industrial-scale open-source benchmarks;
and Section 6 concludes this paper.

\section{Preliminaries} 

This section first states the mission of timing-aware gate-level logic simulation, then talks about the representation and computation of 4-value logic. Finally explains the detail of the delay mechanism in logic simulation.

\subsection{Objective Statement}

The objective of this problem can be stated as follows: Given the design of a netlist, as well as the module functions and delay information of cells that appear in the netlist, calculate the unknown waveforms of all pins for a specific duration according to the available waveforms from some given pins. Further clarification of these items is listed as follows:

\begin{itemize}
    \item{Cell and Netlist: In the netlist, a cell can be regarded as the package of several basic logic gates such as ``AND'' gate and ``NOT'' gate. The pins of internal basic gates compose the pins of a cell. The design of a netlist illustrates the connection relation between the pins of cells.}
    \item{Module Function: In VHDL, each cell is defined as a ``module'', and the module function describes the operation of every internal basic logic gate of a cell. In this paper, the basic logic gates employ 4-value logic.}
    \item{Delay: In reality, a transition triggered by input pins will usually not immediately reflect on the output pin. How to handle the delay between transitions is the most critical hardcore mission to ensure the correct result in simulation tasks, which will be discussed in the following parts of this section. The delay setting of basic logic gates can be very complex, and even two gates with the same logic function in a cell may have different settings.}
    \item{Waveform and Duration: All waveforms appearing in this task start and terminate within a specific duration. Each waveform consists of several consecutive transitions in this duration, where each transition is described by two indicators --- value and time, which denote the value that its signal will change to and the occurrence time of this transition, respectively. The pins with a known waveform are generally primary or pseudo-primary inputs, which compose the initially available waveforms in the task.}
\end{itemize}

\subsection{4-Value Logic}

2-value logic that involves values ``0'' and ``1'' has been well-known in many scientific fields, so it will not be reviewed in this section. 4-value logic is used in VHDL simulation tasks, and it involves four values ``0'', ``1'', ``X'', and ``Z'' that denote the signal of low, high, unknown, and the high impedance state, respectively. The transition of a signal rising from ``0'' is called ``posedge'', while the one falling from ``1' is called ``negedge''.

During the calculation of 4-value logic, value ``Z'' is regarded as ``X'' \cite{hdl}, and value ``X'' shall be considered the possibility to be ``0'' or ``1''. For example, the result of logic ``1 AND X'' depends on the actual value of ``X'' because if ``X'' is ``1'' the result will be ``1'' and ``0'' otherwise, while logic ``1 OR X'' is always ``1'' whatever the value of ``X''.

More details about the truth tables of 4-value ``AND'', ``OR'', and ``XOR'' logic are shown in Table \ref{truetables}.

\renewcommand{\arraystretch}{1.35}

\begin{table}[htbp]
    \caption{Truth Tables of 4-Value ``AND'', ``OR'', and ``XOR'' Logic} 
    \centering
    \setlength{\tabcolsep}{1.7mm}
    \subfloat[AND]{
    \begin{tabular}{c|cccc}
        \hline
                   & \textbf{0} & \textbf{1} & \textbf{X} & \textbf{Z} \\
        \hline
        \textbf{0} & 0          & 0          & 0          & 0          \\
        \textbf{1} & 0          & 1          & X          & X          \\
        \textbf{X} & 0          & X          & X          & X          \\
        \textbf{Z} & 0          & X          & X          & X          \\
        \hline
    \end{tabular}}
    \hfill
    \subfloat[OR]{
    \begin{tabular}{c|cccc}
        \hline
                   & \textbf{0} & \textbf{1} & \textbf{X} & \textbf{Z} \\
        \hline
        \textbf{0} & 0          & 1          & X          & X          \\
        \textbf{1} & 1          & 1          & 1          & 1          \\
        \textbf{X} & X          & 1          & X          & X          \\
        \textbf{Z} & X          & 1          & X          & X          \\
        \hline
    \end{tabular}}
    \hfill
    \subfloat[XOR]{
    \begin{tabular}{c|cccc}
        \hline
                   & \textbf{0} & \textbf{1} & \textbf{X} & \textbf{Z} \\
        \hline
        \textbf{0} & 0          & 1          & X          & X          \\
        \textbf{1} & 1          & 0          & X          & X          \\
        \textbf{X} & X          & X          & X          & X          \\
        \textbf{Z} & X          & X          & X          & X          \\
        \hline
    \end{tabular}}
    \label{truetables}
\end{table}

	


\subsection{Delays of Gates}

The delay from the time an input transition occurs to the time a corresponding output transition occurs is affected by many factors \cite{sdf}. In this task, the following factors should be taken into account for an event:

\begin{itemize}
    \item{The input pin that triggers this transition.}
    \item{The transition edge of the input signal.}
    \item{The transition value of the output signal.}
\end{itemize}

\begin{figure}[htbp]
    \raggedright
    \subfloat[``a'' triggering]{\includegraphics[width=1.05in]{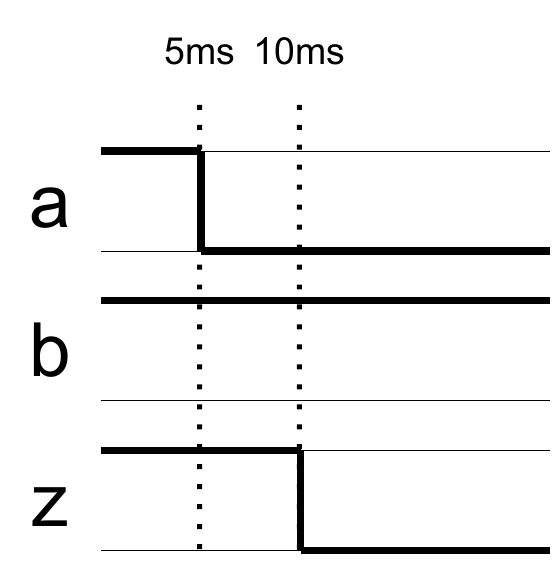}%
        \label{distinct_delay_1}}
    \hfil
    \subfloat[``b'' triggering]{\includegraphics[width=1.05in]{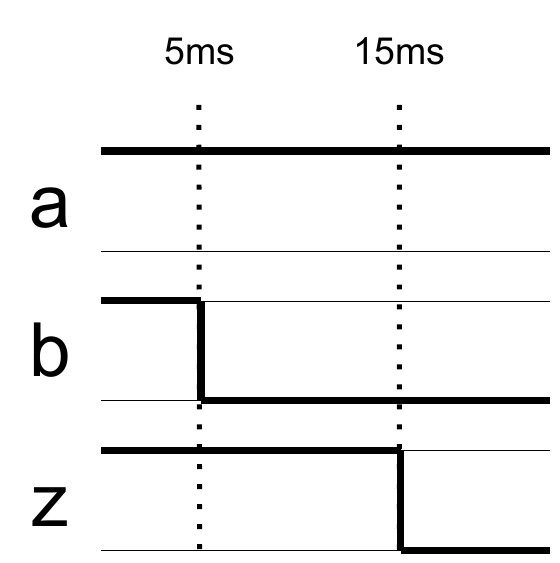}%
        \label{distinct_delay_2}}
    \hfil
    \subfloat[both triggering]{\includegraphics[width=1.05in]{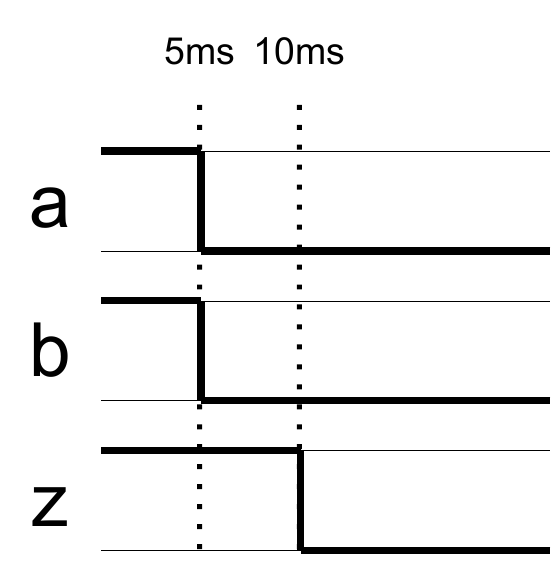}%
        \label{distinct_delay_3}}
    \caption{The cases in which the transitions of distinct input pins occur simultaneously and then result in the same transition occurring in an identical output pin. The minimal delay among these involved pins will be chosen as the final one.}
    \label{distinct_delay}
\end{figure}

Frequently, in a basic logic gate, transitions of distinct input pins may occur simultaneously and contribute to the same transition in an identical output pin. It means several delays will be involved, and the minimal one among them shall be chosen as the final delay of this transition. Fig. \ref{distinct_delay} gives the example of an ``AND'' gate with input pins ``a'', ``b'', and output pin ``z'' to help illustrate this case. The delay of output signal falling triggered by pin ``a'' is 5ps, and the one triggered by pin ``b'' is 10ps, so the final delay will be 5ps when ``a'' and ``b'' fall at the same time.

\begin{figure}[b]
    \raggedright
    \subfloat[``a'' triggering]{\includegraphics[width=1.05in]{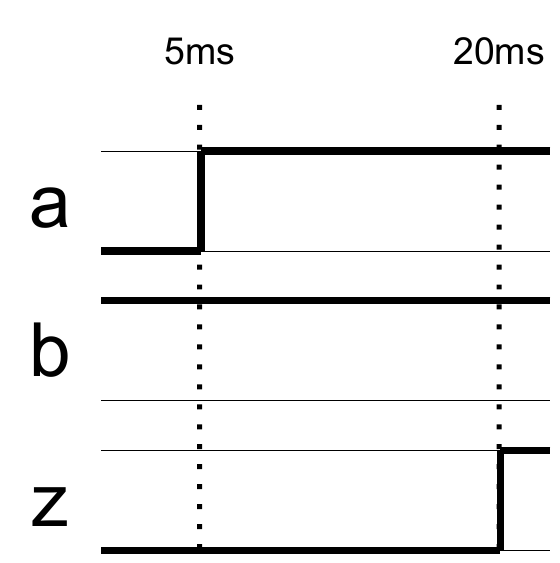}%
        \label{glitch_eaten_1}}
    \hfil
    \subfloat[``b'' triggering]{\includegraphics[width=1.05in]{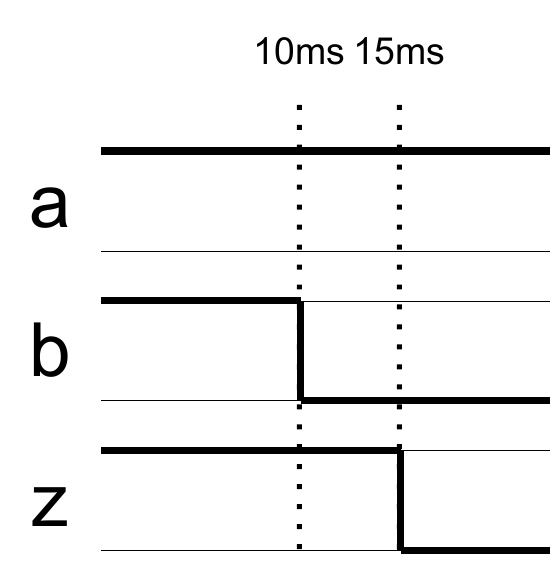}%
        \label{glitch_eaten_2}}
    \hfil
    \subfloat[``glitch eaten'']{\includegraphics[width=1.05in]{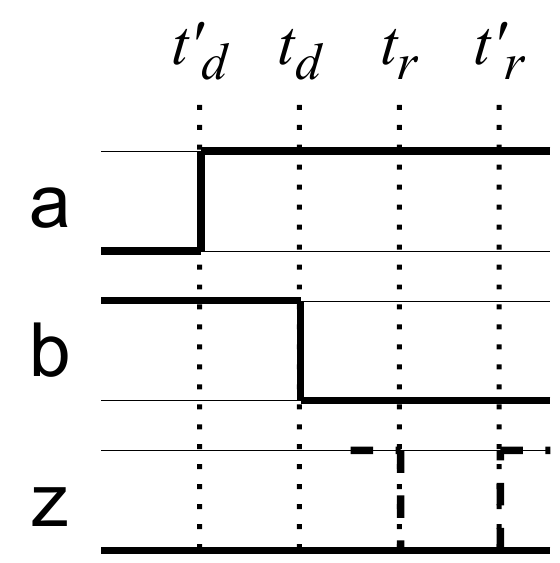}%
        \label{glitch_eaten_3}}
    \caption{An example of ``glitch eaten'', in which a schedule transition is denied and replaced by another one that is determined later and appears earlier. Note that the signal of ``z'' will not be changed since the transition result equals the original signal of pin ``z''.}
    \label{glitch_eaten}
\end{figure}

A transition that has been determined and waiting to appear on the output pin is called a schedule transition. Sometimes one or more schedule transitions may be denied and replaced by a later one. Formally, assuming a transition with determined time $t_d$ and appearance time $t_r$, then it will deny and replace all ones that with determined time $t_d'$ and appearance time $t_r'$ if the following condition is met:
\begin{equation}
    \left\{
    \begin{array}{ll}
        t_d > t_d'\\
        t_r \le t_r'
    \end{array}
    \right.
\end{equation}
This situation is called ``glitch eaten'' in \cite{zhang2020problem}, which may be better known as ``pulse filtering''. Fig. \ref{glitch_eaten} gives the example of an ``AND'' gate with input pins ``a'', ``b'', and output pin ``z'' to help illustrate this case. The delay of the output signal rising triggered by pin ``a'' is 15ps, and the delay of the output signal falling triggered by pin ``b'' is 5ps. So that if ``a'' rises at 5ps and ``b'' falls at 10ps, then the schedule rising of ``z'' at 20ps will be denied and replaced by the falling one at 15ps.


\section{Data Structures Designed for One-Pass Parallelism}

The proposed approach in this paper aims at reducing the synchronization between CPU and GPU during the whole task, particularly to one round of data transfer before and after a one-pass GPU parallelism. In this section, we discuss why the previous approaches cannot afford one-pass parallelism and how our designed data structures bring the opportunity to adopt a new parallel strategy. To achieve this reduction, the raw data involved in the processing must be well organized and transferred to GPU before the parallel method runs. The most important data among them are waveform, delay, and module function.

\begin{figure}[t]
    \centering
    \includegraphics[width=3.3in]{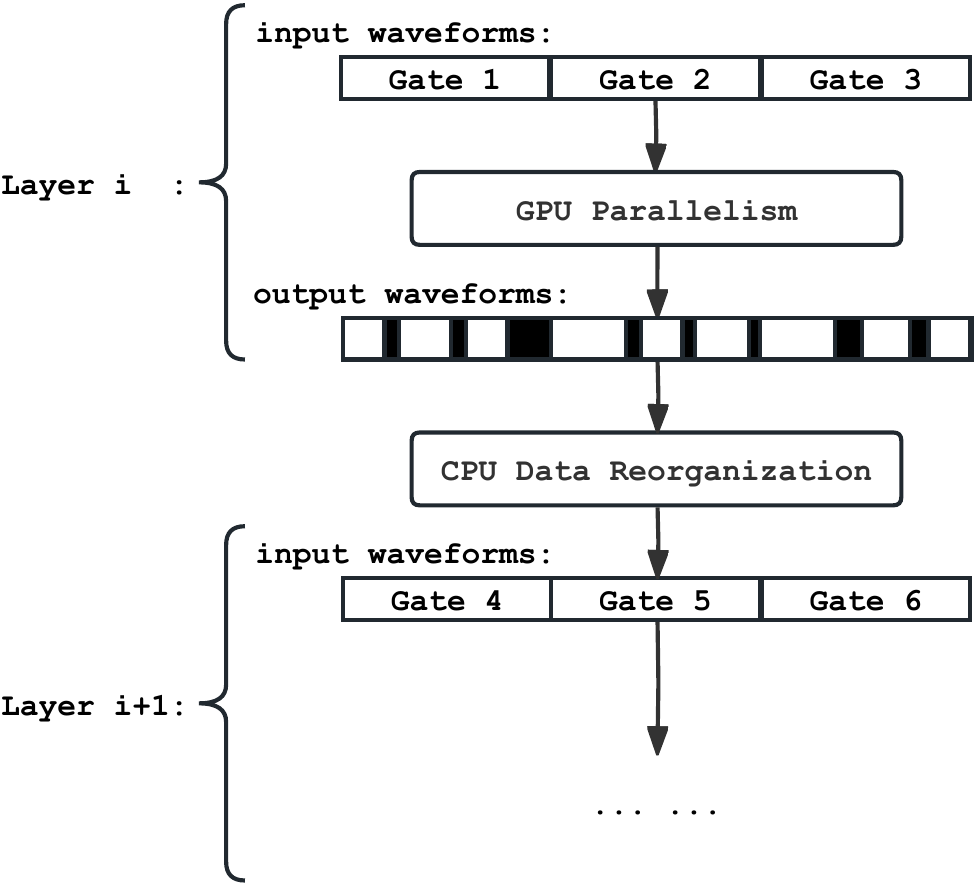}
    \caption{Overview of data structure in event-based parallel simulation. Input and output waveforms are stored in different data structures, and the black blocks in output waveforms indicate the invalid states, such as transitions denied by ``glitch eaten'' and extra allocation preventing memory exceeded. Therefore, the output waveforms cannot be used directly in the following calculation and require a data reorganization by CPU during the synchronization.}
    \label{old_structure}
\end{figure}

\begin{figure}[t]
    \centering
    \includegraphics[width=2.8in]{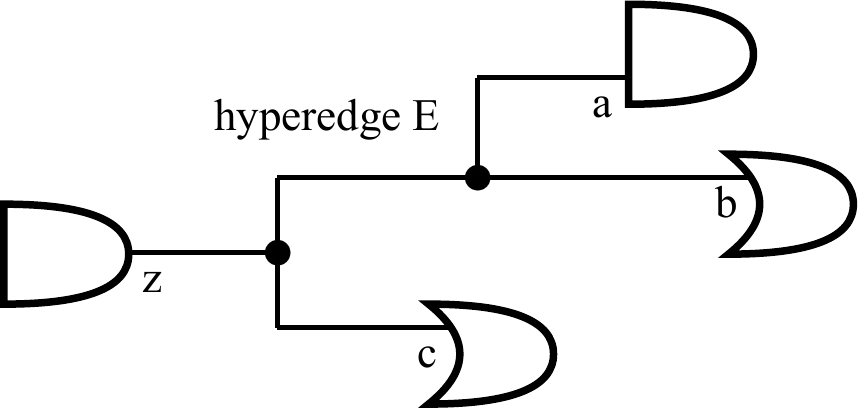}
    \caption{Distribution of an output waveform. Output pin ``z'' is connected with input pins ``a'', ``b'', and ``c'' by a hyperedge ``E''. Storing the waveform on hyperedge ``E'' is enough in theory. Still, the waveform needs to be copied three times to fit the data structures of event-based parallelism illustrated in Fig. \ref{old_structure}.}
    \label{hyperedge}
\end{figure}

In the previous event-based parallel method, the whole simulation task is composed of several parallel layers. Tasks in each layer must be independent because the inputs and outputs of tasks are stored in different data structures \cite{zeng2021accelerate}. As illustrated in Fig. \ref{old_structure}, in event-based parallel methods, all input waveforms of a basic logic gate are stored tightly in a contiguous list, and a list with the same size is allocated to store its output memory because each input transition will trigger up to one output transition, i.e., the size of the output waveform $Size(o)$ has the following upper bound:
\begin{equation}
    \label{layer_estimation}
    Size(o) \leq \sum_{i:\ f(i)=o,\ {\forall}i{\in}I_{layer}}{Size(i)}
\end{equation}
where $I_{layer}$ denotes the set of all input waveforms involved in this parallel layer, $f(i)=o$ indicates $i$ and $o$ is a pair of relative input and output waveforms for a basic logic gate, respectively. CPU data reorganization during synchronization is integral to eliminating invalid states inside the output waveforms and distributing the rest for the inputs of other gates. Besides, an identical output waveform needs to be copied for each input pin during the distribution to hold the event-based parallelism, as explained in Fig. \ref{hyperedge}. Thus the obtained output waveforms cannot be used directly as input waveforms of its successor gates.

To achieve our one-pass parallelism with two rounds of data transfer, the newly designed data structures shall address two purposes:
\begin{itemize}
    \item{Any stored output waveform can be used directly in its successor parallel calculation while the memory is not reallocated during the parallelism, thus getting rid of the synchronization with CPU.}
    \item{Reduce the memory consumption so that massive necessary data for all tasks can be stored in GPU devices}
\end{itemize}
This section only reveals the design of data structures that satisfy our purposes. How our approach leverages these data structures will be discussed in the next section.

\subsection{Waveform}

The waveforms consist of continuous transitions in essence and can be described in various formats such as WLF, VCD \cite{vcd}, and FSDB. In this task, the raw waveform data shall be converted into a matrix at first since the data structure transferred into a GPU device has to be matrixes. For a directly converted two-dimension matrix, the significant elements of each waveform are stored consecutively from the first column of a row, and the extra columns of each row need to be filled for alignment.

\begin{figure}[b]
    \centering
    \subfloat[Conventional CSR structure.]{
        \includegraphics[width=3.1in]{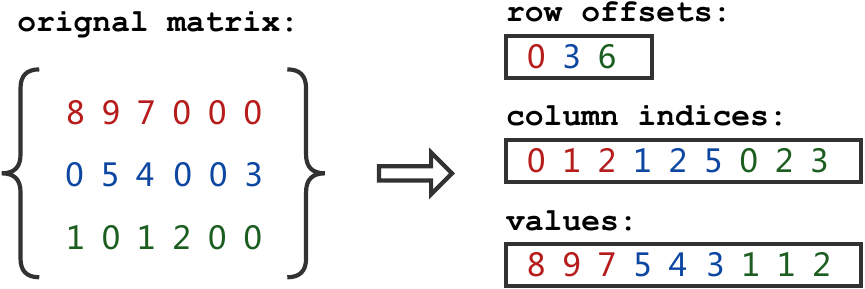}%
        \label{naive_csr}
    }
    \hfil
    \subfloat[Our CSRP structure.]{
        \includegraphics[width=3.1in]{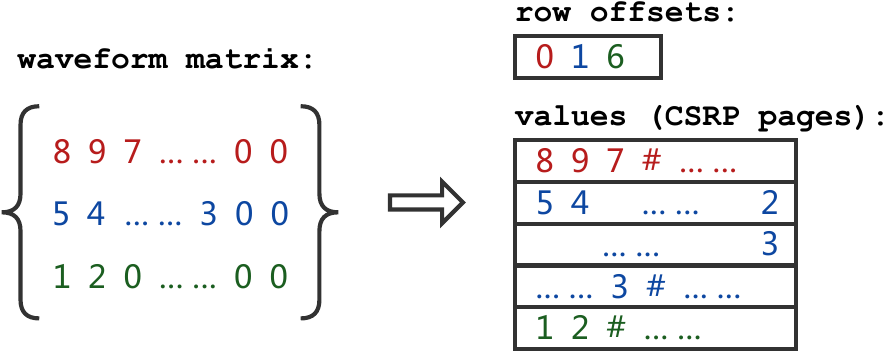}%
        \label{modified_csr_overview}
    }
    \caption{Overview of (a) CSR and (b) our Modified CSRP, where different rows are indicated by distinct colors. In particular, the column indices structure in (b) is omitted  since each waveform is always stored consecutively in a row of a waveform matrix.}
    \label{csr}
\end{figure}

However, the lengths of waveforms among various pins may be significantly different, which will cause serious memory wasting if we directly use a two-dimension matrix to store waveforms. Generally, such a waveform matrix will be sparse and thus can be converted into a compressed representation like Compressed Sparse Row (CSR). The waveform storing structure in Fig. \ref{old_structure} is an implementation of CSR as well. The conventional CSR structure consists of three parts --- row offsets, column indices, and values. Significant elements in the original matrix are stored tightly in values. The $i$th number in row offsets indicates the beginning position in values where the $i$th row of the original matrix is stored, and the $j$th number in the column indices indicates the original column a stored element is located in. Fig. \ref{naive_csr} illustrates an example of the conventional CSR structure. For this problem, transitions are stored consecutively in each row of the original matrix so that the column indices can be omitted.


Unfortunately, another problem emerges when CSR is employed to store the waveforms. The memory of GPU devices must be allocated definitely before the program runs, so the size of values in CSR must be defined before the task. Although the matrix of previously known waveforms can be directly converted into CSR, for those unknown waveforms to be figured out, their precise length can only be obtained after the calculation. If we only have the size of primary and pseudo-primary input waveforms before the task, the size of an output waveform $Size_o$ has the following upper bound:
\begin{equation}
    \label{worst_estimation}
    Size(o) \leq \sum_{i:\ \exists{n}\in{N^+},\ f^n(i)=o,\ \forall{i}\in{I_{netlist}}}{P_i^o * Size(i)}
\end{equation}
where $I_{netlist}$ denotes the set of all primary and pseudo-primary input waveforms in the netlist, $P_i^o$ is the number of distinct paths from $i$ to $o$ in abstract DAG of the netlist, and $f^n(i) = o$ indicates $i$ and $o$ is a pair of relative input and output waveforms pass by $n$ basic logic gate. It can be observed that Eq. \ref{layer_estimation} is a trivial case of Eq. \ref{worst_estimation} when $n=1$ because event-based methods introduce the assistance of CPU during synchronization to estimate and reallocate the size of output waveforms for single parallel layers. Obviously, we cannot employ this upper bound in one-pass parallelism, or it may even incautiously cause the estimated allocation of memory to exceed the capacity of GPU devices rather than merely waste. Moreover, the complete order of calculation in parallelism is hard to guarantee, which further restricts the design of recyclable memory.


\begin{figure}[t]
    \centering
    \includegraphics[width=2.8in]{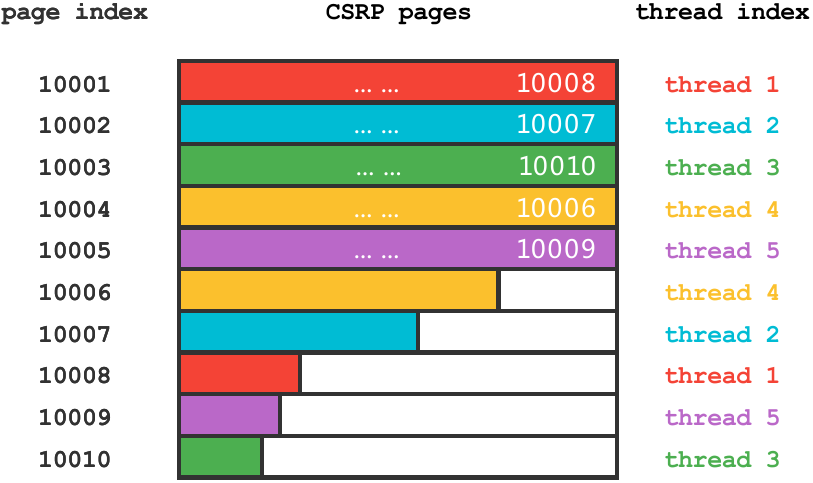}
    \caption{The reason for introducing next-page pointers. While multi GPU threads are running, the physical next empty page is likely to be employed by another thread before a thread finishes its work on its current using page, thus making a waveform to be stored in inconsecutive pages. Hence, it is necessary to use the last value of a page to point to the logical next page.}
    \label{modified_csr_threads}
\end{figure}

To tackle this issue, we propose a modified structure called Compressed Sparse Row in Pages (CSRP), which is illustrated in Fig. \ref{modified_csr_overview}. The proposed CSRP structure can dynamically store new-generated uncertain-scale output waveforms in GPU devices. Inspired by the idea of the page table structure in the field of Operating System, the contiguous list of values in the conventional CSR is modified to a matrix, where each row of this matrix is called a page. Let $pagelen$ be the length of a page in CSRP then each page stores $pagelen-1$ elements. The last value of a page is called a next-page pointer which indicates the index of the next page that stores the same waveform in CSRP. Using a next-page pointer to indicate the index of its next page is necessary because the obtained waveform may hardly be stored in consecutive pages while multi GPU threads are running, as is illustrated in Fig. \ref{modified_csr_threads}. After the storing of a row from the original matrix is finished at a certain page, a terminate value (represented as ``\#'' in this example) will be set following the last store element at that page. The terminate value is not forbidden to be set at the last position of the page. Assuming there are $k$ waveforms in total, then the waste of memory $M_w$ has the following upper bound:
\begin{equation}
    \label{memory_waste}
    M_w \le pagelen * k * M_t
\end{equation}
where $M_t$ denotes the necessary memory allocated to store a single transition. Notice that despite the illustration being divided into stages, raw waveform data are converted directly into CSRP in practice without an intermediate waveform matrix. A newly obtained waveform to be stored will dynamically ask for a new page once its current distributed memory is inadequate. This way, we maintain the given and unknown waveforms in the same structure without remaining any duplication and allow GPU threads to easily find and traverse the stored output waveforms as input.

\subsection{Delay}

A five-dimension matrix is used to store the delay information. These five dimensions denote the index of cells, the index of input pins, the index of output pins, the transition edge of the input signal, and the transition value of the output signal, respectively. Each dimension is a discrete variable that can be identified by finite numbers. Some invalid elements are natural to occur in this matrix, e.g., in a cell, an input pin may have no relation with several output pins. The value of these invalid elements will be filled as $inf$, an inconsequent big number denoting infinite.

The edge of a transition is recorded at the corresponding timestamp during the calculation, i.e., we know whether an input signal is changed at any timestamp. To figure out the final delay of a transition, we need just traverse the 2nd dimension of the delay matrix to check all changed input signals at a specific timestamp and take the minimal delay value among them, while the indices of other dimensions are definitive and unique for each output transition. We do not need to maintain the relationship between input signals and output signals using extra resources since the delay value has been set as $inf$ if an input signal has no relation to the output signal, which means the transition of irrelevant input signals will not contribute to the result.

\subsection{Module Function}

The module function is not necessary to be implemented for each cell because many cells are derived from the same standard cell template. Their association is similar to the one between object and abstract class.

To establish the module function library for all cells, our approach only needs to implement the logic functions of all basic logic gates and combine them according to the requirement of any standard cell template appearing in the netlist. This way, our approach also succeeds in handling User Defined Primitives (UDP). We record the computing logic of the corresponding pins for the basic logic gates in the standard cell template. These data are read-only and thus can be shared by many threads at the same time.

\section{Waveform-Based GPU Accelerated Simulation Method}

This section first analyses the disadvantage of the previous method applied in DAG, and then proposes a waveform-based parallel method and explains its operation.

\subsection{Parallelism Management}

\begin{figure}[t]
    \centering
    \subfloat[DAG indicating the topological order of the tasks using partitioning strategy based on topological sorting. The value alongside each node indicates the time cost of the corresponding task. The total time cost is 60ps since it has to add up the time cost of the slowest task in every layer, i.e., task $A$, $E$, and $I$.]
    {\hspace{0.35in}\includegraphics[width=2.5in]{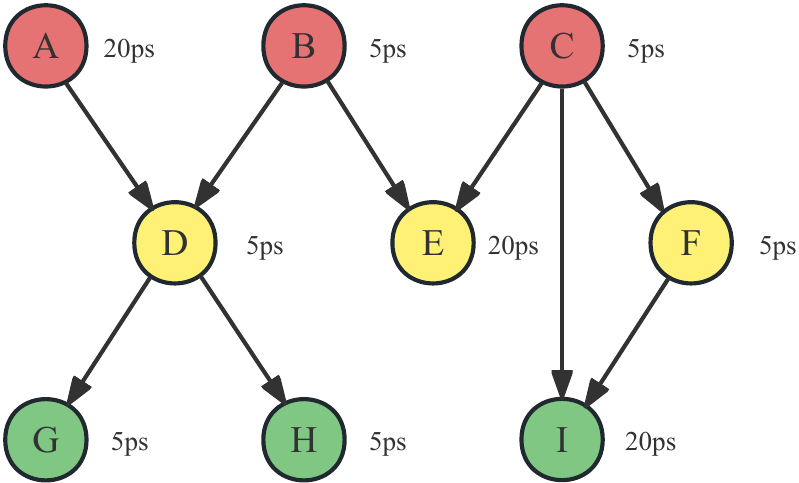}\hspace{0.35in}%
        \label{topology_conventional}}
    \hfill
    \subfloat[Timeline of tasks in partitioning strategy, omitting the time cost of synchronization. All tasks in the same layer cost 20ps in total. The partitioning strategy based on topological sorting fails to parallelize $A$ and $F$, which are independent tasks in different layers and should be parallelized.]
    {\hspace{0.35in}\includegraphics[width=2.5in]{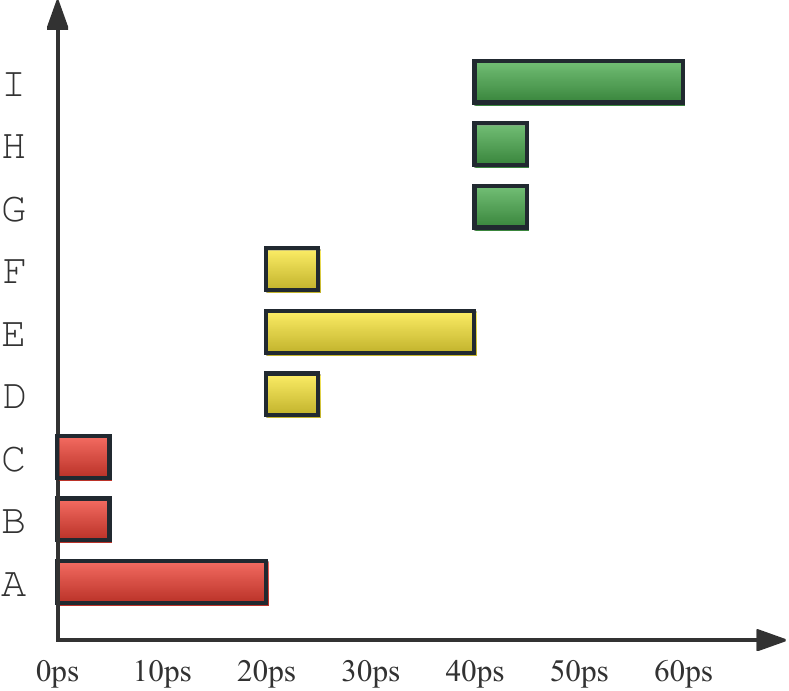}\hspace{0.35in}%
        \label{timeline_conventional}}
    \caption{Illustration of the partitioning strategy based on topological sorting and its defect. Different colors  are used to indicate layers.}
    \label{partition}
\end{figure}

As the abstract graph of the circuit is DAG, it is obvious to come up with a strategy that partitions and parallelizes the task based on their dependency or topological relation. And this partitioning strategy is exactly widely researched in many existing works about various parallel simulation tasks such as \cite{sen2010parallel}, \cite{zhang2011logic}, \cite{zeng2021accelerate}, \cite{zhu2017leveraging}, and \cite{li2020quantum}. In this strategy, the original DAG is partitioned by parallel layers, each with several parallelizable tasks. However, every two adjacent parallel layers demand a synchronization, which forces all tasks in the next layer, regardless of dependent or not, to wait for the completion of the slowest task in the previous layer, thus failing to utilize the computing resources of GPU threads entirely. Besides, each parallel layer generally requires a round of exclusive data transfer between CPU and GPU during the synchronization to exchange current relevant data supporting calculation.

Fig. \ref{topology_conventional} provides an example to illustrate this issue. The graph is partitioned into three layers using topological sorting. Only the cells in the same layer will be processed in parallel. For each layer, the start of any parallel task has to wait for the slowest 20ps-cost task in the previous layer to complete, which results in a final time cost of 60ps in total, as shown in Fig. \ref{timeline_conventional}. Strictly, task F is allowed to begin as long as task C is completed without waiting for the finish of task A. This result suggests that the effectiveness of the conventional partitioning strategy is affected by the concrete partition result. Many works have made efforts to investigate appropriate methods of arranging the parallel tasks in different coarse grains, like dividing task A into sub-task A' and A'' to reduce the time wasted on the next layer waiting for the previous layer. However, such methods certainly extend the critical path in the DAG, which increases the time consumption on extra rounds of data transfer and results in the consideration of a new trade-off problem. Moreover, as an inherent problem in partitioning strategies, this issue can never be perfectly solved because a partition can only reflect the dependency between tasks in macro. Benefitting from the rapidly growing memory of modern developing GPU devices, we can now invent new ideas that try to skip this classical problem.

\begin{algorithm}[t]
    \SetAlgoLined
    \caption{Missions in Parallel inside Each GPU Thread}
    \label{gpu_thread_flow}

    \KwIn {Known waveforms. Information of cells.}
    \KwOut {All waveforms.}
    \While {True}
    {
        $done$ = True \\
        \ForEach {cell $i$ in arranged cells}
        {
            \If {checkCellProcessed($i$) is True}
            {
                \textbf{continue}
            }
            $done$ = False \\
            $flag$ = True \\
            \ForEach {input waveform $j$ of cell $i$}
            {
                \If {checkWaveformKnown($j$) is False}
                {
                    $flag$ = False
                }
            }
            \If {$flag$ is True}
            {
                processCell($i$) \\
                \ForEach {output waveform $k$ of cell $i$}
                {
                    storeWaveform($k$)
                }
            }
        }
        \If {$done$ is True}
        {
            \textbf{break}
        }
    }
\end{algorithm}

\begin{figure}[t]
    \centering
    \subfloat[Step-wise execution of our non-partitioning strategy regarding the time slots. In the figure, the colors white, blue, pink, and red respectively indicate (1) tasks that are not started, (2) tasks under processing, (3) tasks that are under processing and will be completed at the end of this time slot, and (4) completed tasks.  
    ]
    {\hspace{0.05in}\includegraphics[width=3.1in]{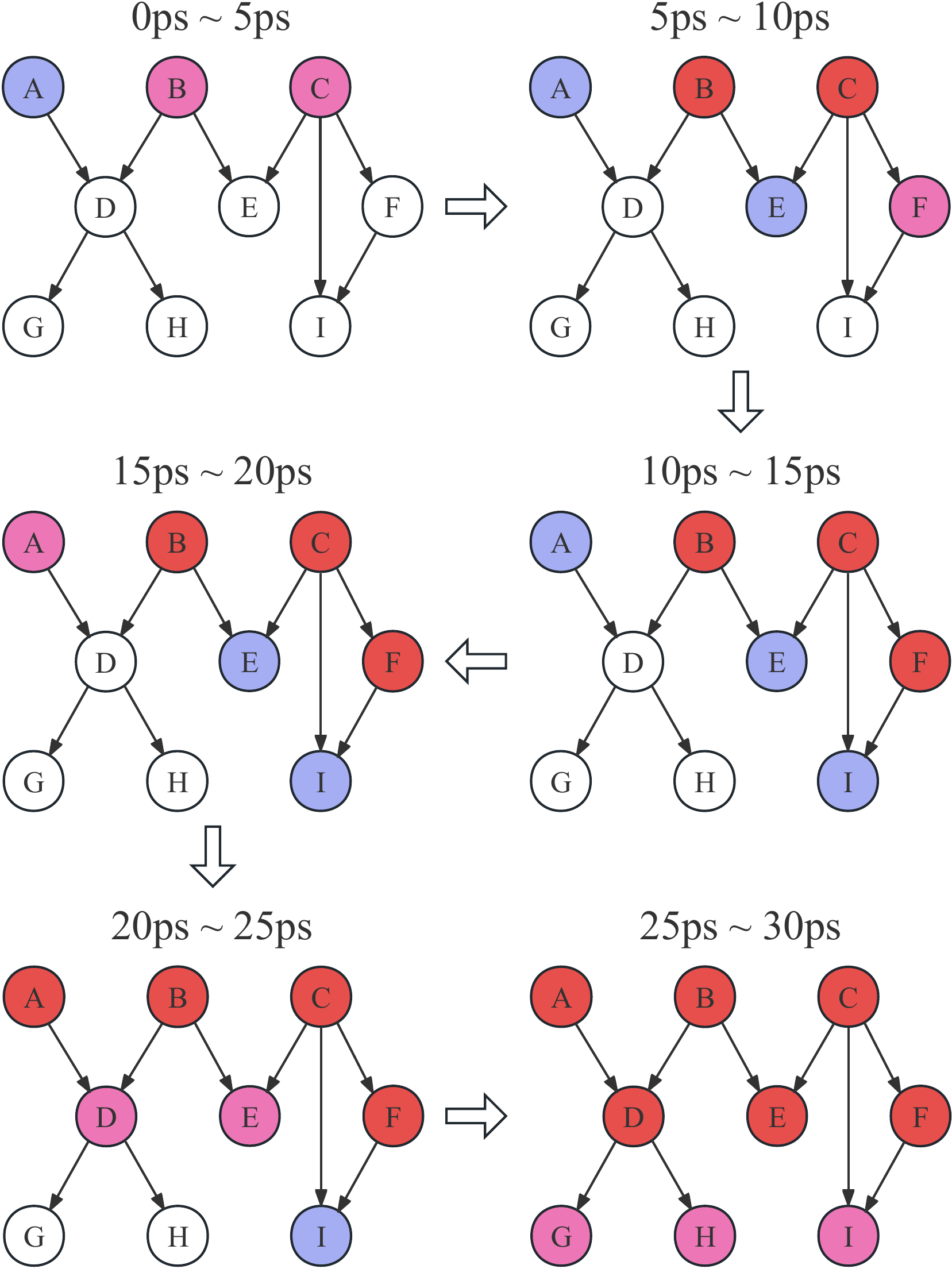}\hspace{0.05in}%
        \label{topology_novel}}
    \hfill
    \subfloat[Ideal timeline of tasks in our non-partitioning strategy. A task can be immediately triggered once all its predecessor tasks are completed, even if they are in different topological layers. The total time cost of all tasks is reduced to 30ps.]
    {\hspace{0.35in}\includegraphics[width=2.5in]{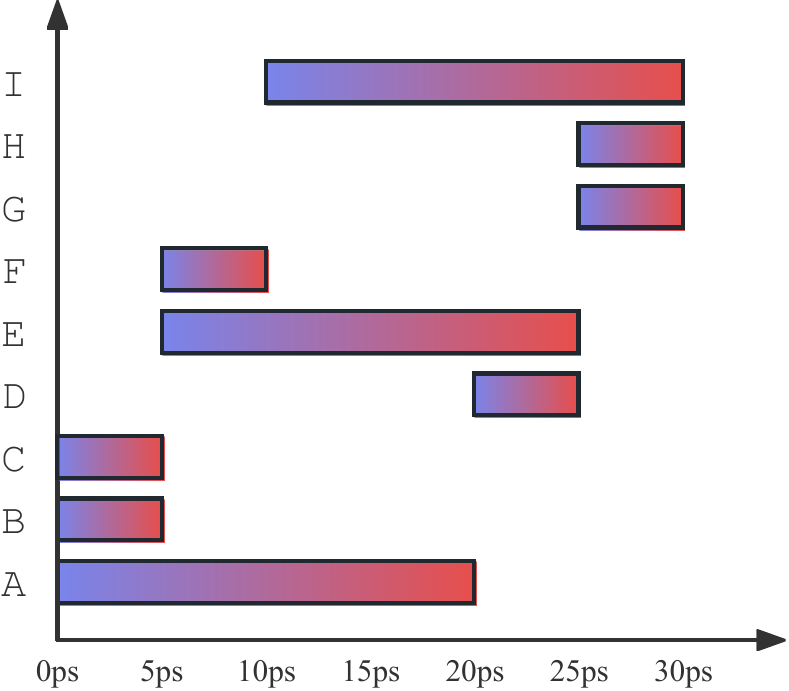}\hspace{0.35in}%
        \label{timeline_novel}}
    \caption{Illustration of our non-partitioning strategy based on the direct dependency between tasks. Different colors are used to indicate the processing stage of a task.}
    \label{non-partition}
\end{figure}

Ample memory supplied by modern GPU devices generates the possibility of transferring all relevant data through a single procedure and then employing the GPU parallel algorithm to check whether the start condition of a task is met. Furthermore, if each task can begin immediately after all its predecessor tasks are complete, the time cost of the case in Fig. \ref{partition} will be decreased to 30ps. Fig. \ref{non-partition} shows the step-by-step operation of such a non-partitioning strategy applied in these tasks.

The mission of simulation on combinational circuits can be regarded essentially as waveforms to waveforms based on the above discussion. Hence, a waveform-based GPU parallel strategy is proposed in this paper. We arrange the cells to be processed in the netlist equally on all GPU threads. For example, assuming there are $n$ cells and $m$ threads, the number of cells arranged on each thread is $n/m$ or $n/m+1$. As for the case that the number of cells is less than the number of GPU threads, each thread will serve a cell and process it immediately while the condition meets. A cell is ready to be processed only if the thread confirms that all waveforms of its input pins are known, or it will be called a locked cell otherwise. Since the netlist is a DAG and thus the necessary input waveforms of a cell can be figured out before calculating its output waveforms, the proposed waveform-based method holds. Leveraging designed structures, we can achieve one-pass parallelism for simulation tasks within the provided memory of current advanced GPU devices. If the GPU memory cannot store all the necessary information for an extremely huge task, the method can also introduce some easy means to divide the task into a few parts. Note that such a division is different in essence from the layer partition in existing methods since we do not ask that the arranged tasks in GPU devices must be independent of each other.

The overall processing flow of a GPU thread is designed as Algorithm \ref{gpu_thread_flow}. Initially, only the given waveforms can be obtained. Every running thread will conduct an endless loop to check the cells arranged on it for whether they can be processed or its processing is finished. When all arranged cells on a certain thread finish processing, the thread will be terminated, and the whole GPU parallelism will finish as long as all its threads are terminated.

\begin{algorithm}[t]
    \SetAlgoLined
    \caption{Calculation for Output Waveforms of an Unlocked Cell}
    \label{calculate_signal}

    \KwIn {Input waveforms $W_{in}$ = \{$W_{in}^1$, $W_{in}^2$, ..., $W_{in}^I$\}. $W_{in}^i$ = \{($w_{i1}.v, w_{i1}.t$), ..., ($w_{iJ_i}.v, w_{iJ_i}.t$)\}, where $w_{ij}.v$ and $w_{ij}.t$ denotes the value and timestamp of the $j$th transition in the $i$th waveform, respectively; Delay information of this cell $Delay$, i.e., the first dimension of the delay matrix is omitted here; Static module function library that consists of various combinations of basic gate logic functions for all templates.}
    \KwOut {Output waveforms $W_{out}$ = \{$W_{out}^1$, $W_{out}^2$, ..., $W_{out}^K$\}.}
    $currentSignals$ = \{X, X, ..., X\} \\
    $earliestIndex$ = \{1, 1, ..., 1\} \\
    $earliestTimestamp$ = \{$w_{11}.t$, $w_{21}.t$, ..., $w_{I1}.t$\} \\
    \While{True}
    {
        $t_{earliest}$ = $inf$ \\
        \ForEach{$t$ in $earliestTimestamp$}
        {
            $t_{earliest}$ = $min(t_{earliest}, t)$
        }
        \If{$t_{earliest}$ == $inf$}
        {
            \textbf{break}
        }
        \ForEach{input waveform $W_{in}^i$}
        {
            $j$ = $earliestIndex_i$ \\
            \If{$w_{ij}.t$ == $t_{earliest}$}
            {
                update $w_{ij}.v$ to $currentSignals$ \\
                record the transition edge $e_i$ of $W_{in}^i$\\
                \uIf{$j$ == $J_i$}
                {
                    $earliestTimestamp_i$ = $inf$
                }
                \Else
                {
                    $earliestIndex_i$ = $j$ + 1 \\
                    $earliestTimestamp_i$ = $w_{ij+1}.t$
                }
            }
        }
        $newSignals$ = calculateSignals($currentSignals$) \\
        \ForEach{output signal value $o_k.v$}
        {
            \If{$o_k.v$ is changed}
            {
                $del_k$ = $inf$ \\
                \ForEach{input waveform $W_{in}^i$}
                {
                    $del_k$ = $min(del_k, Delay[i][k][e_i][o_k.v])$
                }
                $o_k.t$ = $t_{earliest}$ + $del_k$ \\
                addSignalChange($o_k.v$, $o_k.t$, $W_{out}^k$)
            }
        }
        $currentSignals$ = $newSignals$
    }
\end{algorithm}

\subsection{Storing of Outputs}

Once a cell can be processed, its output waveforms will be calculated according to its module functions and delay setting, and meanwhile, update to the CSRP structure. After that, the updated waveform will be marked as known and available. Notice that while some waveforms are marked to be available by a thread, the other threads are still checking whether their arranged cells can be processed. Once the input waveforms of a locked cell are all calculated, the cell will be unlocked and processed immediately. Algorithm \ref{calculate_signal}, i.e., function \textit{processCell()} in line 15 of Algorithm \ref{gpu_thread_flow}, shows how to calculate the output waveforms for a cell whose input waveforms are available.

\begin{figure}[b]
    \centering
    \includegraphics[width=3.2in]{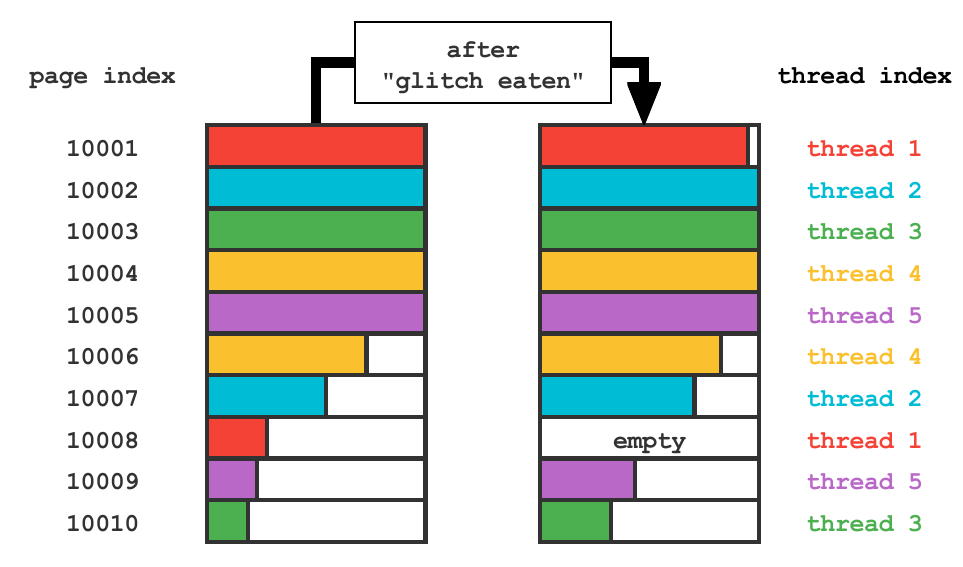}
    \caption{Illustration for the issue of storing calculated transitions directly and immediately in the CSRP structure. As the backtrace modification on CSRP structure caused by ``glitch eaten'' may cross pages, in this example, the backtrace crosses from the 10008th page to the 10001st page. It makes the 10008th page empty, but the iterator maintained by the \textit{atomic-add} function has been greater than 10008. Thus the 10008th page cannot be reused.}
    \label{empty_page}
\end{figure}

The management of empty pages in CSRP is based on the \textit{atomic-add} function \cite{api} in CUDA. We employ this function to maintain a monotonic increasing iterator points to the index of the first remaining empty page. Once a thread requires a new page, the empty page currently pointed by the iterator will be selected, and then the value of the pointed index will be plus by 1. The atomic function can avoid the involvement of multi threads in this process, which ensures the result of parallel computation is correct. The given waveforms and the unknown waveforms to be calculated are stored in the same CSRP structure. Each thread reads the input waveforms of a cell from it and stores the obtained output waveforms in it for other threads to read. When every thread finishes its work, this CSRP structure will be transferred to the CPU as the output of simulation tasks.

\begin{figure}[t]
    \centering
    \includegraphics[width=2in]{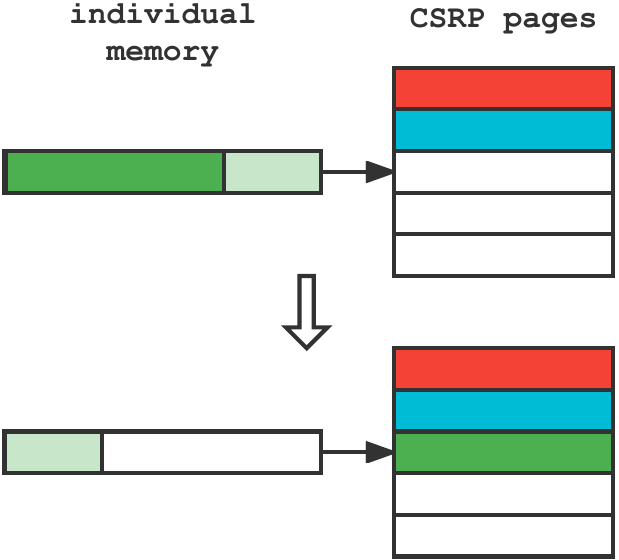}
    \caption{Communication between individual memory and the modified CSRP structure. Once the individual memory is used up, the prefix transitions of one-page size are stored in an empty page in the CSRP structure, whereas the remains are moved to the front.}
    \label{individual_memory}
\end{figure}

During storing the calculated waveforms, our algorithm needs to handle invalid states caused by the ``glitch eaten'' issue, which means that some schedule transitions will be denied and replaced by a later triggered one. Only after that can we use these output waveforms directly in the successor calculation. The ``glitch eaten'' will not happen more than once at the same time in 2-value logic because there are only two kinds of schedule transitions that can replace each other, i.e., once a schedule ``0'' denies and replaces a schedule ``1'', it will be equal to the current signal or the previous schedule signal, which is also ``0'', and vice versa. But ``glitch eaten'' may happen recursively in 4-value logic, which makes it hard to ensure how many schedule transitions will be involved. If we directly and immediately update the schedule transitions to CSRP, the backtrace modification of transitions caused by ``glitch eaten'' may cross pages, which causes an empty page appears in the previously distributed memory. However, the pages are distributed by the \textit{atomic-add} function in order, so the appeared empty page cannot be reused and thus resulting in an undesirable waste of memory, as shown in Fig. \ref{empty_page}.

We attempt to eliminate this problem by allocating an individual memory of 1.5x page size for each thread. The calculated transitions will temporarily be stored in the individual memory. When a memory is full, its prefix transitions of one-page size will be transferred into an empty page in CSRP, and the remains will be moved to the front, which is shown in Fig. \ref{individual_memory}. After the transfer, the index of the used new page will be filled at the end of the last page used by that thread as the next-page pointer, and then the value of the new page iterator will be plus 1 by the \textit{atomic-add} function. In theory, this trick is acceptable only if the backtrace of a waveform never crosses half a page. Still, it has been proven reasonable in experiments because the capacity of half a page will certainly be far greater than the number of recursive ``glitch eaten'' in practice.

\section{Experiment}

\renewcommand\arraystretch{1.35}

\begin{table*}[htbp]
	\caption{Statistics on Benchmarks}
	\centering
	\setlength{\tabcolsep}{1.2mm}
	\label{statistics}
	\begin{tabular}{clrrrrrrr}
		\hline
		      & \multicolumn{1}{c}{design}  & \multicolumn{1}{c}{No. of cells} & \multicolumn{1}{c}{duration (ps)} & \multicolumn{1}{c}{No. of basic gates} & \multicolumn{1}{c}{No. of given waveforms} & \multicolumn{1}{c}{given waveform length} & \multicolumn{1}{c}{WCV} \\
		\hline
		{1 } & RISCV\_DefaultConfig\_random & 129,993 & 19,990,001    & 296,977           & 28,562          & 28,615,983      & 0.04     \\
		{2 } & RISCV\_TinyConfig\_median    & 41,661  & 2,539,935,010 & 93,645            & 11,686          & 15,904,296      & 7.42     \\
		{3 } & RISCV\_TinyConfig\_random    & 41,661  & 99,990,001    & 93,645            & 11,686          & 58,309,799      & 0.04     \\
		{4 } & NV\_NVDLA\_partition\_c      & 59,830  & 2,972,036,001 & 154,329           & 32,000          & 44,672,321      & 17.10    \\
		{5 } & NV\_NVDLA\_partition\_m      & 11,648  & 2,972,036,001 & 27,259            & 2,098           & 34,817,491      & 5.50     \\
		{6 } & NV\_NVDLA\_partition\_o      & 99,481  & 2,972,036,001 & 247,075           & 38,539          & 13,118,860      & 54.54    \\
		\hline
	\end{tabular}
\end{table*}

In this paper, we conduct experiments on several industrial-scale open-source re-simulation benchmarks derived from \cite{riscv} and \cite{nvdla}. The code is written in Python and runs in NVIDIA GeForce GTX 1080 with merely no more than 8GB memory available. The compilation of parallel Python code is supported by Numba for CUDA GPUs \cite{numba}. The statistics on the benchmarks are shown in Table \ref{statistics}, where ``given waveform length'' indicates the sum of all given waveforms' lengths and ``WCV'' indicates the coefficient of variation (CV) about the set of all given waveforms’ lengths. WCV is calculated as follows:
\begin{equation}
    \label{coefficient_variation}
    WCV = \dfrac{\sigma_{lengths}}
                {\bar{X}_{lengths}}
\end{equation}
where $\sigma_{lengths}$ and $\bar{X}_{lengths}$ denote the standard deviation and the average about the set of all given waveforms' lengths, respectively. This metric will be further validated in the following analysis of experiments. It is introduced to estimate the probability of short tasks waiting for long tasks caused by the conventional partitioning strategy, witnessing the time cost of a task highly depends on the length of its input waveforms as is evident in Algorithm \ref{calculate_signal}. These statistics are to conclude the meaningful information associated with this task so it may not be coincident with the official description of designs, e.g., we only count those cells that need to be processed.

The length of a CSRP page in the experiments of these experiments is set as 256. In order to fit the NumPy \cite{numpy} support in Numba for CUDA GPUs, the memory of a page element is set as 9 Bytes, which is the sum of NumPy.int64 (for time) and NumPy.int8 (for value). For the parallel settings, we arranged 128 threads for a block and 40 blocks for a grid. The shared memory of blocks fails to be utilized since the required memory of all individual memory in a block reaches 128 $*$ 256 $*$ 1.5 $*$ 9 $=$ 432KB, which exceeds the limitation of $48$KB per block. Although executing a sharp decrease in the size of CSRP pages can achieve the utilization of shared memory, it may not benefit the performance of the parallelism because more storing operations from individual memory to CSRP will be required simultaneously. And our storing operation is based on the \textit{atomic-add} function, which may cause it to be more time-consuming when reducing the size of CSRP pages.

\begin{table}[t]
    \caption{Comparison of Speedup with Normalization}
    \centering
    \setlength{\tabcolsep}{2.85mm}
    \label{speedup}
    \begin{tabular}{crrr}
        \hline
              & \multicolumn{1}{c}{The 1st place of \cite{zhang2020problem}}  & \multicolumn{1}{c}{Paper \cite{zeng2021accelerate}}  & \multicolumn{1}{c}{This paper}  \\
        \hline
        { 1 } & 13.35                                                         & \textbf{13.78}                                       & 6.33                            \\
        { 2 } & 10.38                                                         & 7.00                                                 & \textbf{11.51}                  \\
        { 3 } & 7.33                                                          & \textbf{10.78}                                       & 8.72                            \\
        { 4 } & 2.41                                                          & 3.52                                                 & \textbf{5.57}                   \\
        { 5 } & 13.33                                                         & \textbf{15.05}                                       & 5.19                            \\
        { 6 } & 1.75                                                          & 1.91                                                 & \textbf{10.05}                  \\
        Avg.  & 6.21                                                          & 6.87                                                 & \textbf{7.55}                   \\
        \hline
    \end{tabular}
\end{table}

The metric of performance in the experiments of this paper is measured with speedup, which is calculated as follows:
\begin{equation}
    \label{measure}
    \begin{array}{ll}
        & Speedup = \dfrac{t_{serial}}
                          {t_{parallel}} \\
        & Average\ Speedup = \sqrt[n]{\prod\limits_{i=1}^n{Speedup_i}}
    \end{array}
\end{equation}
where $t_{serial}$ is the runtime of our self-implement event-based serial approach and $t_{parallel}$ is the runtime of our proposed GPU accelerated approach. We compare our results with the 1st place of problem \cite{zhang2020problem} in ICCAD contest 2020 and another existing work \cite{zeng2021accelerate}, whose results are developed by C++ while ours is developed by Python. We investigate the classical $n$-queen problem and large-scale pixel traverse problem, and the result suggests that the naive Python version runs dozens of times slower than the C++ version, which also meets the observation of our serial code and the given C baseline. Hence, comparing the runtime with the C++ baseline makes no sense. For fairness, we normalize our speedup to compare it with theirs since their speedup is based on the baseline with a 16-core CPU parallelism. We refer to CPU utilization to evaluate the speedup of the 16-core CPU parallelism and minify our speedup in the comparison. This evaluation is based on the achieved best CPU utilization aggregated average across all 16 available CPUs in problem \cite{zhang2020problem}, which is presented by the statistics in the follow-up study \cite{zeng2021accelerate}.

\begin{table}[t]
    \caption{Comparison of GPU Utilization}
    \centering
    \setlength{\tabcolsep}{2.5mm}
    \label{gpu_utilization}
    \begin{tabular}{ccc}
        \hline
                                                 & {NV\_NVDLA\_m}  & {NV\_NVDLA\_o}  \\
        \hline
        The 1st place of \cite{zhang2020problem} & 41.80\%         & 12.70\%         \\
        The 2nd place of \cite{zhang2020problem} & 12.80\%         & 2.55\%          \\
        The 3rd place of \cite{zhang2020problem} & 21.80\%         & 30.00\%         \\
        Paper \cite{zeng2021accelerate}          & 40.50\%         & no result       \\
        This paper                               & 99.50\%         & 87.71\%         \\
        \hline
    \end{tabular}
\end{table}

The comparison indicates that our approach gains a significant edge over the others on benchmarks 4 and 6. The speedup on these two benchmarks is increased by 58.2\% and 426.1\% compared with the best existing result, respectively. It is worth noting that these two benchmarks possess the third and second most cells, respectively, and also the longest duration. More cells appearing in the design makes the task more computation-intensive and also means the conventional partitioning strategy, such as topological sorting, may arrange more parallelizable cells into different layers. On the other hand, the longer duration means the time costs of tasks are likely to be different, i.e., some tasks will be significantly longer than others, thus forcing many tasks to waste much more time waiting for the longest task in their previous parallel layer. The probability of such an undesirable waste of time is also reflected in the ``WCV'' summarized in Table \ref{statistics}, where benchmarks 6 and 4 possess the highest and the second-highest ``WCV'' among all, respectively. As well as more data involved in the calculation demands more synchronization between CPU and GPU devices, which further reduces the efficiency of the whole parallelism. Our approach tackles this issue by employing GPU code to control the processing of dependent tasks and transferring all necessary data into the GPU device before the parallelism begins, thus outperforming the conventional partitioning strategy and event-based method.

The speedup on other benchmarks further provides evidence for the analysis that our approach performs better in situations with more cells and longer duration. Benchmarks 2 and 3 possess identical designs, and the performance of our approach on benchmark 2 is better than benchmark 3 because the former last for a longer duration. In contrast, our approach performs much worse in benchmarks 1 and 5. According to the statistics and comparison on benchmarks, the most relative feature to our performance is ``WCV'', which can infer that the most significant improvement of our approach is successfully tackling the issue when the previous approaches applied in tasks with different time costs. Nevertheless, our architecture actually still retains the extensibility of incorporating local event-based parallelism to preserve its performance for parallel tasks with similar time costs.

\begin{table}[t]
    \caption{GPU Memory Usage}
    \centering
    \setlength{\tabcolsep}{3.65mm}
    \label{memory_usage}
    \begin{tabular}{lc}
        \hline
        \multicolumn{1}{c}{design}  & Memory Usage (MB) \\
        \hline
        RISCV\_DefaultConfig\_random & 6,612             \\
        RISCV\_TinyConfig\_median    & 4,095             \\
        RISCV\_TinyConfig\_random    & 4,041             \\
        NV\_NVDLA\_partition\_c      & 3,049             \\
        NV\_NVDLA\_partition\_m      & 5,975             \\
        NV\_NVDLA\_partition\_o      & 3,571             \\
        \hline
    \end{tabular}
\end{table}

As for the GPU utilization measured by command \textit{nvidia-smi}, we conduct experiments for comparison between ours and all existing approaches in other literature on benchmarks 5 and 6 in table \ref{gpu_utilization}, demonstrating that our parallel strategy attains a high score in parallelism. Provided that a thread is possibly either busy working or waiting during the procession, despite the  GPU utilization  metric  for the parallel strategy might compromise in some cases in the comparison,  it is demonstrated that our approach is with a better performance regarding parallelism on average.

Table \ref{memory_usage} summarizes the actual memory usage of these benchmarks by accurately calculating the size of the sufficient allocated memory for GPU devices. Note that memory usage of a benchmark depends on the intrinsic information of the design as well as the scale of waveforms changing in the duration. The results suggest that our approach is expected to process the design tenfold to the used benchmarks, i.e., with millions of logic gates and proportionately larger scale of waveforms, through one-pass parallelism in a single advanced GPU with 80GB memory, and can potentially succeed in a larger task using integrated GPU workstations or future productions.

\section{Conclusion}

In this paper, we proposed a one-pass waveform-based GPU accelerated parallel approach against large-scale combinational circuit simulation based on a high-performance computing system with modern GPUs. To this end, we devised a set of relevant data structures to reduce the communication consumption between CPU and GPU, achieving one-pass parallelism that needs only one round of data transfer. With the designed structures, we proposed a waveform-based parallel strategy that overcomes the defect of the previously existing partitioning strategy,  achieving a higher GPU utilization rate. Experimental results demonstrated that our approach could achieve a better average performance compared to SOTA. The performance gap is more significant when  the benchmarks are with more cells and longer duration, and is particularly better against the benchmarks with a bigger ``WCV'' which indicates greater different time costs during parallelism, manifesting that our approach compares favorably in computation-intensive/large-scale tasks than the existing approaches.

\ifCLASSOPTIONcompsoc
  \section*{Acknowledgments}
\else
  \section*{Acknowledgment}
\fi

This work is supported by the National  Science Foundation of China (Nos. 12271098, 61772005).

\ifCLASSOPTIONcaptionsoff
  \newpage
\fi



%

%

\begin{IEEEbiography}[{\includegraphics[width=1in,height=1.25in,clip,keepaspectratio]{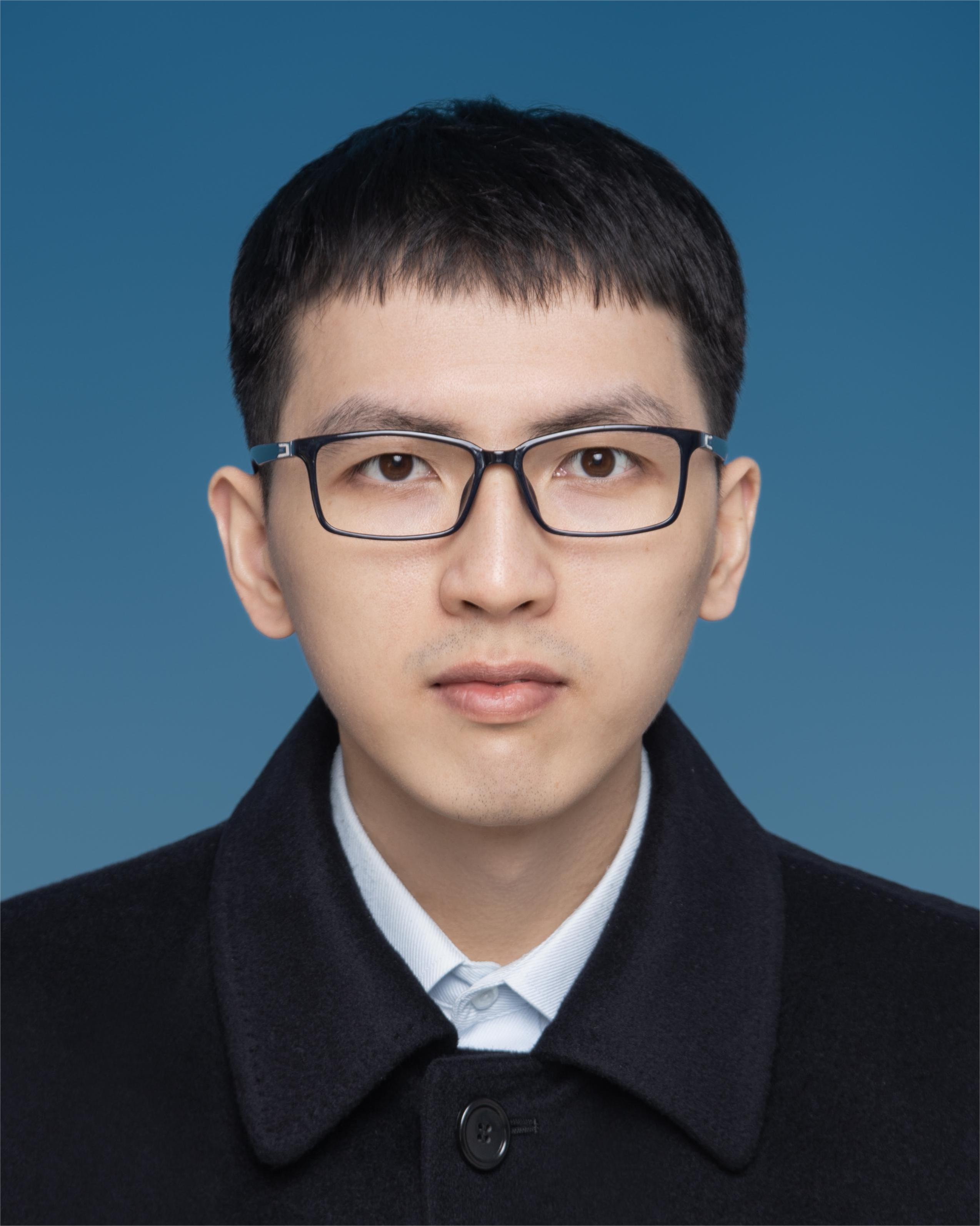}}]{Weijie Fang}
received B.S. and M.S. degrees in Computer Science from Fuzhou University of China in 2017 and 2020, respectively. He is currently pursuing his Ph.D. degree in the School of Mathematics and Statistics at Fuzhou University. His research interests include efficient algorithm design, electronic design automation, and intelligent decision technology. His research interests include optimization problems in high-performance computing systems, computer-aided design on VLSI, etc.
\end{IEEEbiography}
\vspace{-5pt}

\begin{IEEEbiography}[{\includegraphics[width=1in,height=1.25in,clip,keepaspectratio]{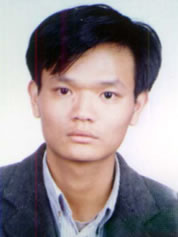}}]{Yanggeng Fu}
received Ph.D. degree from Fuzhou University of China in 2013. He is currently a full professor at the College of Computer and Data Science/College of Software at Fuzhou University. He has published over 50 academic papers. His research interests include data mining, machine learning, and intelligent decision support systems.
\end{IEEEbiography}
\vspace{-5pt}

\begin{IEEEbiography}[{\includegraphics[width=1in,height=1.25in,clip,keepaspectratio]{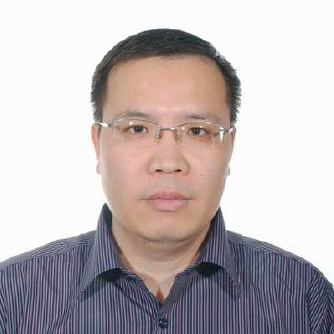}}]{Jiaquan Gao}
received Ph.D. degree in Computer Science from the Institute of Software, Chinese Academy of Sciences in 2002. He is currently a professor with the School of Computer and Electronic Information at Nanjing Normal University in Nanjing, China. He respectively worked as a visiting scholar at McGill University, Canada, from September 2007 to September 2008 and at Georgia Institute of Technology, US, from December 2011 to May 2012. His current research interests include high-performance computing (HPC), parallel algorithms on heterogeneous platforms for solving linear/nonlinear systems, and computational intelligence.
\end{IEEEbiography}
\vspace{-5pt}

\begin{IEEEbiography}[{\includegraphics[width=1in,height=1.25in,clip,keepaspectratio]{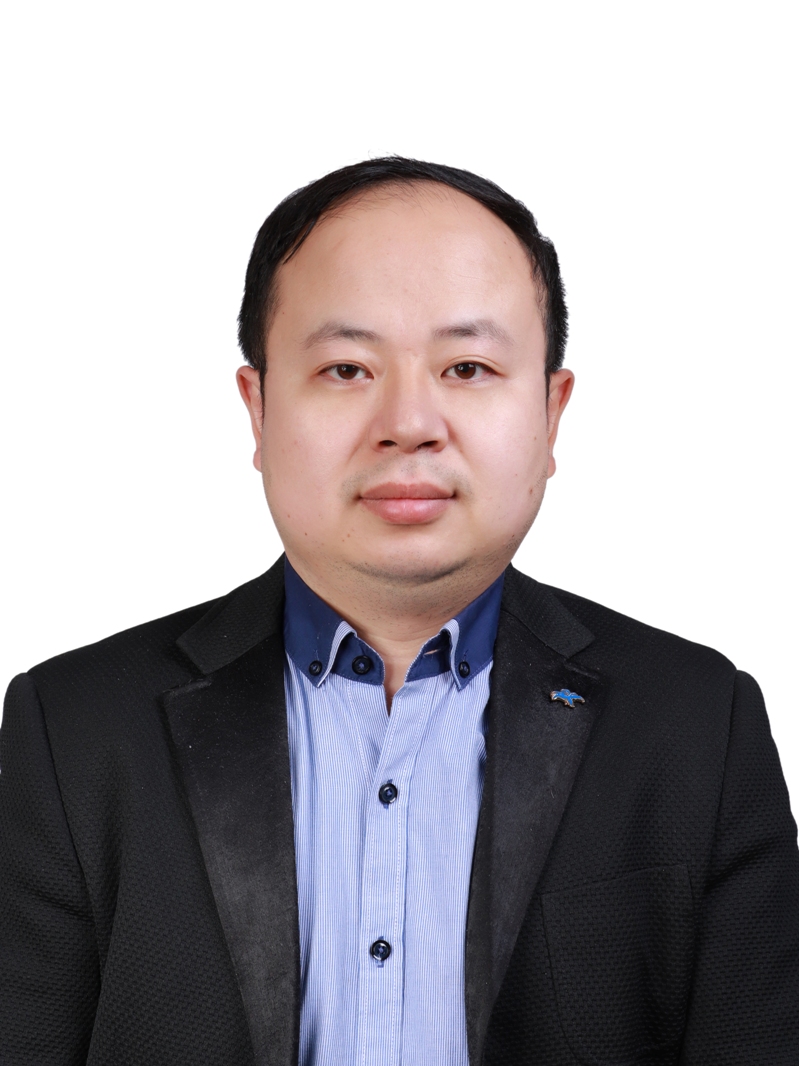}}]{Longkun Guo}
received B.S. and Ph.D. degrees in Computer Science from the University of Science and Technology of China (USTC) in 2005 and 2011, respectively. He is currently a full professor at Fuzhou University. His major research interests include efficient algorithm design and computational complexity analysis, particularly for optimization problems in high-performance computing systems and networks, VLSI, etc. He has published more than 100 academic papers in reputable journals/conferences such as IEEE TMC, IEEE TC, Algorithmica, IEEE TPDS, IEEE ICDCS, IJCAI, and ACM SPAA.
\end{IEEEbiography}
\vspace{-5pt}

\begin{IEEEbiography}[{\includegraphics[width=1in,height=1.25in,clip,keepaspectratio]{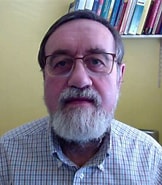}}]{Gregory Gutin}
received Ph.D. degree in mathematics from Tel Aviv University (supervised by Noga Alon) in 1993. Since September 2000, he has been a Full Professor of computer science at the Royal Holloway University of London. He has over 260 publications, which were cited over 11 300 times. His major research interests include graph theory, algorithms and complexity, combinatorial optimization, information security, and theoretical economics. He received the Royal Society Wolfson Research Merit Award in 2014. He Was also elected to Academia Europaea in 2017.
\end{IEEEbiography}
\vspace{-5pt}

\begin{IEEEbiography}[{\includegraphics[width=1in,height=1.25in,clip,keepaspectratio]{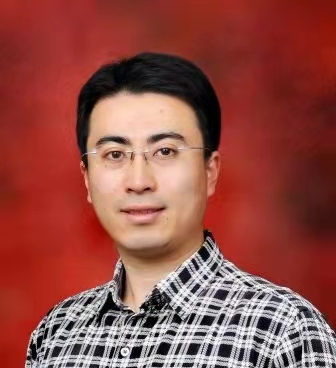}}]{Xiaoyan Zhang}
received the first Ph.D. degree in Applied Mathematics from Nankai University in 2006 and received the second Ph.D. degree in Computer Science from Twente University in 2014. He is currently a full professor at School of Mathematical Science and Institute of Mathematics, Nanjing Normal University. He has published more than 60 academic papers in reputable journals such as SIAM J. Computing, SIAM J. Scientific Computing, SIAM J. Discrete Math, J. Graph Theory, and IEEE Transactions on Information Theory. His major research interest includes efficient algorithm design and computational complexity analysis, particularly for combinatorial optimization, graph algorithms and networks, VLSI, etc.
\end{IEEEbiography}





\end{document}